\documentclass [10pt, a4paper] {article}
\usepackage{color,makeidx}
\usepackage[T1]{fontenc}
\usepackage{graphicx}
\usepackage{enumerate}
\usepackage{amsmath}
\usepackage{amsthm}
\usepackage{amssymb}
\usepackage{pdfsync}
\usepackage{hyperref}
\usepackage[all]{xy}
\makeindex
\newtheorem{theorem}{{\bf{Theorem}}}[section]
\newtheorem{lemma}[theorem]{{\bf{Lemma}}}
\newtheorem{proposition}[theorem]{{\bf{Proposition}}}

\theoremstyle{remark}
\newtheorem{remark}[theorem]{{\bf{Remark}}}

\newcommand{\bzeta}{\boldsymbol\zeta}
\newcommand{\blambda}{\boldsymbol\lambda}

\numberwithin{equation}{section}
\begin{document}
\title{Hydrodynamic Limit for a Hamiltonian system with Boundary
  Conditions and Conservative Noise}
\author{Nadine Even\\Stefano Olla}

\date{\today}

\maketitle

\abstract{We study the hyperbolic scaling limit for a chain of N
  coupled anharmonic oscillators. The chain is attached to a point on
  the left and there is a force (tension) $\tau$ acting on the right.
In order to provide  good ergodic properties to the system, we
perturb the Hamiltonian dynamics with random local exchanges of
velocities between the particles, so that momentum and energy are
locally conserved. We prove that in the macroscopic limit the
distributions of the elongation, momentum and energy, converge
to the solution of the Euler system of equations, in the smooth regime.}

\section{{Introduction}\label{intro}}

The aim of this paper is to study the hydrodynamic limit for a
non-equilibrium system subject to an exterior time dependent force at
the boundary.  
We consider the most simple mechanical model with non-linear
interaction, i.e. a one dimensional chain of $N$ anharmonic oscillators. 
The left side is attached to a fixed point, while on the right side is acting
a force $\tau$ (tension). For each value of $\tau$ there is a family
of equilibrium (Gibbs) measures parametrized by the temperature (and by the
tension $\tau$). It turns out that these Gibbs 
measures can be written as a product.

We are interested in the macroscopic non-equilibrium behaviour of 
this system as $N$ tends to infinity, after rescaling space and time
with $N$ in the same way (\emph{hyperbolic scaling}). 
We also consider situations in which the tension $\tau$ depends slowly on
time, such that it changes in the macroscopic time scale.
In this way we can also take the system originally at
equilibrium at a certain tension $\tau_0$ and push out of equilibrium
by changing the exterior tension.

The goal is to prove that the 3 conserved quantities (elongation,
momentum and energy) satisfy in the limit an autonomous closed set of 
hyperbolic equations given by the Euler system. 

We approach this problem by using the \emph{relative entropy method}
(cf. \cite{yau}) as already done in \cite{ovy} for a system of
interacting particles moving in $\mathbb R^3$ (\emph{gaz dynamics}). 

The relative entropy method permits, in general, to obtain such
hydrodynamic limit if the system satisfy certain conditions:
\begin{enumerate}[A)]
\item\label{erg}
 The dynamic should be \emph{ergodic} in the sense that the only 
conserved quantities that \emph{survive} the limit as $N\to \infty$
are those we are looking for the macroscopic autonomous behavior (in
this case elongation, momentum and energy). More precisely, the only
stationary measure for the infinite system, with finite local entropy,
are given by the Gibbs measures. 
\item\label{smooth} The macroscopic equations have smooth solutions.
\item\label{curr} Microscopic currents of the conserved quantities should be
  bounded by the local energy of the system.
\end{enumerate}
 
We do not know any deterministic hamiltonian system that satisfy
condition \ref{erg}, and this is a major challenging open
problem in statistical mechanics. Stochastic perturbation of the
dynamics that conserves energy and momentum can give such ergodic
property and have been used in \cite{ovy} (cf. also \cite{ls,flo,ffl}). 
We use here a simpler
stochastic mechanism than in \cite{ovy}: at random independent
exponential times we exchange the momentum of nearest neighbor
particles, as if they were performing an elastic collision. 
Under this stochastic dynamics, every stationary measure has the
property to be excheangeable in the velocity coordinates, and this is
sufficient to characterize it as a convex combination of Gibbs
measures (cf. \cite{ffl} and \cite{stefano}). 

About condition \ref{smooth}, it is well known that nonlinear
hyperbolic equations in general develop shocks also starting from
smooth initial condition. Characterization and uniqueness of weak
solutions in presence of shock is a challenging problem in the theory
of hyperbolic equations. We expect that a shock will increase
 the thermodynamic entropy associated to the profiles of the conserved
 quantities. 

The relative entropy method compares the
microscopic Gibbs entropy production (associated to the probability
distribution of the system at a given time) with the macroscopic 
(thermodynamic) entropy production. If no shocks are present both
entropy productions are small. 
The presence of the boundary force changes a bit this balance, since
one should take into account the (macroscopic) change of entropy due
to the work performed by the force. It turns out that the right choice
of the boundary conditions in the macroscopic equation compensate this
large entropy production, keeping the time derivative of the relative
entropy small. 
It would be interesting to prove similar cancellation of entropy
productions when this is caused by shocks, as it would allow to prove
the hydrodynamical limit in these cases, and provide a microscopic
derivation of irreversible thermodynamic \textbf{adiabatic} transformations,
between thermodynaic equilibrium states that
increase the thermodynamic entropy. Recent efforts in this
direction use  different methods (cf. \cite{fritz}).
Similar results on isothermal transformation are mathematically
easier (cf. \cite{ollaiso}).

About condition \ref{curr}, it created a problem in \cite{ovy}:
in the usual gaz dynamics the energy current has the convecting term
cubic in the velocities, while energy is quadratic. This was fixed in
\cite{ovy} by modifying the kinetic energy of the model: if the
kinetic energy grows linearly as a function of the velocity, the energy
current will grow also linearly. 
Since we work here in lagrangian coordinates, our energy current does
not have the cubic convecting term. This allows us to work with the
usual quadratic kinetic energy.  
 
\thanks{This paper has been partially supported by European Advanced
  Grant Macroscopic Laws and Dynamical Systems (MALADY) (ERC AdG
  246953) and by a grant of the MAThematics Center Heidelberg
  (MATCH).}

\section{{The Model and the Main Theorem}\label{MODEL}}

We will study a system of $N+1$ coupled oscillators in one dimension.
Each particle has the same mass that we set equal to 1.
The position of atom $i$ is denoted by $q_i\in \mathbb R$, while its
momentum is denoted by $p_i\in\mathbb R$. 
 Thus the configuration space is $(\mathbb R\times \mathbb R)^{N+1}$. 
We assume that an extra particle $0$ to be attached to a fixed point
and does not move, i.e. $(q_0,p_0)\equiv(0,0)$,
 while on particle $N$ we apply a force $\tau(t)$ depending on time.
Observe that only the particle 0 is constrained to not move, and that
$q_i$ can assume also negative values. 
\begin{center}
\begin{figure}[h]
\includegraphics[width=0.8\textwidth]{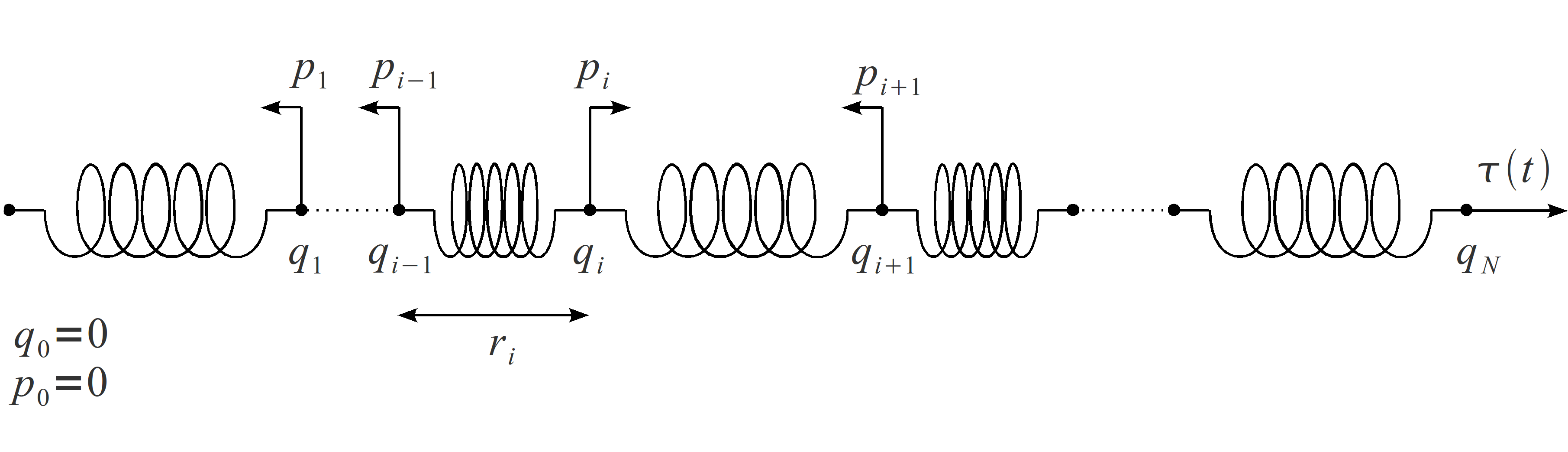}\\	
\end{figure}
\end{center}
Denote by  ${\bf q} :=(q_0,\dots,q_N)$ and ${\bf p}
:=(p_0,\dots,p_N)$. The interaction between two particles $i$ and
$i-1$ will be described by the potential energy $V(q_i-q_{i-1})$ of an
anharmonic spring relying the particles. We assume   $V$ to be a
positive smooth function that grows quadratically at infinity,
i.e. there exists strictly positive constant $C_-, C_+, b$ such that
for any $r\in \mathbb R$:  
\begin{equation}
  \label{V1}
  C_- r^2 \le V(r) \le C_+ r^2 + b .
\end{equation}
Energy is defined by the following Hamiltonian:
\begin{eqnarray*}
\mathcal H_N(\bf q ,\bf p ):
&=&\sum_{i=1}^{N}  \left( \frac{p_i^2}2 + V(q_{i}-q_{i-1}) \right).
\end{eqnarray*}
Since we focus on a nearest neighbor interaction, we may define the distance between particles by \index{{\large{CHAPTER 2:}}!$r_i$}
$$
r_i=q_{i}-q_{i-1}, \qquad i=1,\dots,N.
$$ 
We define the energy of particle $i\in\{1,\dots,N\}$ as
$$
e_i:=\frac{p_i^2}{2}+ V(r_i)
$$
so that $\mathcal H_N({\bf {r} ,\bf {p}} )=\sum_{i=1}^{N}e_i$,
where ${\bf r} :=(r_1,\dots,r_{N})$.

Given a smooth function $\tau(s)$ that represents the force applied to
the particle $N$ at the macroscopic time $s$, the dynamics of the
system is determined 
by the generator 
\begin{equation}\label{noisygen}
 N G_N^{\tau(t)}:= N L^{\tau(t)}_N + N \gamma S_N.
\end{equation}
Here the Liouville operator  $L^{\tau}_N$ is given by
\begin{eqnarray}\label{liouville}
\nonumber L^{\tau}_N
 &=&\nonumber\sum_{i=1}^{N}(p_{i}-p_{i-1})\frac{\partial}{\partial
   r_i}+\sum_{i=1}^{N-1}\left(V^{\prime}(r_{i+1})-V^{\prime}(r_{i})\right)
 \frac{\partial}{\partial p_i}\\ 
 &&+\left(\tau -V^{\prime}(r_N)\right)\frac{\partial}{\partial p_N},
 \end{eqnarray}
where  we used the fact that $p_0\equiv0$. Notice that the time scale
in the tension is chosen such that it changes smoothly on the
macroscopic scale.

The symmetric operator $S_N$ is the generator of the stochastic part
of the dynamics that exchange at random time velocities of nearest
neighbor particles.  
  For any  smooth function $f$, we define the operator $\Upsilon_{i,i+1}$ by 
  \begin{equation}\label{upsilon}
  \Upsilon_{i,i+1}=\frac12\left(f\left(\mathbf  r,\mathbf p^{i,i+1}\right)-f\left(\mathbf r, \mathbf p\right)\right)
  \end{equation}
   where $\mathbf p^{i,i+1}\in{\mathbb R^N}$ is defined from $\mathbf
   p\in\mathbb R^{N}$ by exchanging the coordinates $p_j$ and
   $p_{j+1}$ 
  \begin{equation*}
   p_j^{i,i+1}=\left\{
  \begin{array}{lcl}
  p_j&{\text{if}}&j\neq i,i+1\\
  p_{i+1}&{\text{if}}&j= i\\
  p_{i}&{\text{if}}&j= i+1
  \end{array}\right..
  \end{equation*}
Then $S_N$ is defined through
  \begin{eqnarray}\label{sym}
  S_Nf(\mathbf r,\mathbf p)&:=&\sum_{i=1}^{N-1}\left(f\left(\mathbf r,\mathbf p^{i,i+1}\right)-f(\mathbf r,\mathbf p)\right)\\
  &=&-\sum_{i=1}^{N-1}\Upsilon^2_{i,i+1}f(\mathbf r,\mathbf p)=
  2\sum_{i=1}^{N-1}\Upsilon_{i,i+1}f(\mathbf r,\mathbf p), 
  \end{eqnarray}
With this choice of the noise,
 the  three balanced quantities, i.e. locally conserved,
  are given by $r_i, p_i, e_i$.
  
We define $\boldsymbol\zeta(r,p) = (r, p, -e(r,p))^T\in \mathbb R^2 \times
\mathbb R_-$, and  
 the partition function $Z$ on  
$\mathbb R^2 \times R_+$ by  
\index{{\large{CHAPTER 2:}}!$Z(\bold\lambda)$} 
  $$
  Z({\boldsymbol\lambda}) = Z(\lambda_1, \lambda_2,\lambda_3) :=
  \int_{\mathbb R^2}e^{\boldsymbol\lambda\cdot \boldsymbol\zeta(r,p)} \;dr dp.    
  $$
and the canonical Gibbs function as its logarithm: 
\begin{equation}\label{freeenergy}
 \Theta(\boldsymbol\lambda):=\log Z\left(\boldsymbol\lambda\right),
 \end{equation}
 By the condition imposed on $V$, this function is always finite.

For $\boldsymbol\zeta\,\in \mathbb R^2\times \mathbb R_-$  we define
$\Phi:\mathbb R^2\times \mathbb R_-\rightarrow\mathbb R$ by the
Legendre transform of  the canonical Gibbs function  
 \begin{equation}\label{entropy}
 \Phi(\boldsymbol\zeta):=
 \sup_{\boldsymbol\eta \in \mathbb R^2\times\mathbb R_+}
\left\{\boldsymbol\eta
  \cdot\boldsymbol\zeta-\Theta(\boldsymbol\eta)\right\}. 
 \end{equation}
So that the inverse is 
 \begin{equation}\label{freenergy2}
 \Theta(\boldsymbol\eta):=
 \sup_{\boldsymbol\zeta \in \mathbb R^2\times\mathbb R_-}
\left\{\boldsymbol\eta
  \cdot\boldsymbol\zeta(r,p) -\Phi(\boldsymbol\zeta)\right\}. 
 \end{equation}

 We denote by ${\boldsymbol\lambda}(\tilde{\mathfrak u})$ and 
 $\tilde{\mathfrak u}(\boldsymbol \lambda):=(\mathfrak r,\mathfrak p,-E)^T$ 
the corresponding convex conjugate variable, that satisfy
\begin{equation}\label{dual}
\boldsymbol\lambda=D\Phi(\tilde{\mathfrak u})\quad{\text{and}}\quad
\tilde{\mathfrak u}=D\Theta(\boldsymbol\lambda) ,
\end{equation}
where the operator $D$ is defined by
 \begin{equation}\label{D}
 Df(\mathbf a):=\left(\frac{\partial f}{\partial a_1},\frac{\partial f}{\partial a_2},\frac{\partial f}{\partial a_3}\right)
 \end{equation}
for any $C^1$ function $f:\mathbb R^3\rightarrow\mathbb R$ and $\mathbf a:=(a_1,a_2,a_3)\in\mathbb R^3$.

On the one particle state space $\mathbb R^2$ we define a family of
probability measure
\begin{equation}
  \label{eq:marginals}
  \nu_{\boldsymbol \lambda} (dr, dp) = e^{\boldsymbol\lambda
  \cdot\boldsymbol\zeta(r,p) - \Theta(\boldsymbol \lambda)} dr dp
\end{equation}
Observe that
\begin{equation}\label{expectation}
\nonumber E_{\nu_{\boldsymbol\lambda}}[\boldsymbol\zeta(r,p)] 
= \tilde{\mathfrak u}
\end{equation}
so we can identify $\tilde{\mathfrak u} =(\mathfrak r, \mathfrak p, -E)^T$ as
respectively the average distance, velocity and (negative) energy.
We also define the \emph{internal energy}  $\mathfrak e = E- \mathfrak p^2/2$
.
We have the relations
 $$
E_{\nu_{\boldsymbol\lambda}}(p^2) - \mathfrak p^2 = \lambda_3^{-1} :=
\beta^{-1},\qquad 
P(\mathfrak r,\mathfrak e):= E_{\nu_{\boldsymbol\lambda}}[V^{\prime}(r)] = 
\frac {\lambda_1}{\lambda_3} :=\tau
$$
that identify $\beta^{-1}$ as temperature and $\tau$ as tension.
This thermodynamic terminology is justified by observing that, for
constant $\tau$ in the dynamics, and any $\beta>0$, 
with the choice $\boldsymbol\lambda = (\beta\tau, 0 , \beta)$
the family of product measures given by:  
 $$
 \nu_{(\tau\beta,0,\beta)}^N(d{\bf r},d{\bf
   p}) = \prod_{i=1}^N\nu_{(\tau\beta,0,\beta)}(dr_i,dp_i), \qquad 
 \beta\in\mathbb R^+ 
  $$
is stationary for the dynamics. 
These are
 the grand canonical Gibbs measures at an average  temperature
$\beta^{-1}$, pressure $\tau$ and velocity $0$.  

  In what follows  we need also Gibbs measure with average velocity
  different from $0$, and we will use the following notation:
  \begin{equation*}
  \nu_{\boldsymbol\lambda}^N:=
  \prod_{i=1}^N 
  e^{\boldsymbol\lambda \cdot \boldsymbol\zeta_i -\Theta(\boldsymbol\lambda)}
  dr_idp_i := g^N_{\boldsymbol\lambda}(\mathbf r,\mathbf p) 
   d\mathbf rd\bold p,
  \end{equation*}
  
  where
$\bzeta_i:=(\zeta_{i,1},\zeta_{i,2},\zeta_{i,3})^T :=(r_i,p_i,-e_i)^T$.

In a similar way we may introduce the local Gibbs measures: For
any continuous \emph{profile} $\tilde{\mathfrak u}(x)$, $x\in [0,1]$,
we have correspondingly a \emph{profile} of parameters
$\boldsymbol\lambda(x)$, and we define the inhomogeneous product
measure 
\begin{equation*}
\nu^N_{\boldsymbol\lambda(\cdot)}:=
\prod_{i=1}^N 
e^{\boldsymbol\lambda(i/N) \cdot \bzeta_i -\Theta(\boldsymbol\lambda(i/N))}
  dr_idp_i ,
\end{equation*}
that we call \emph{Local Gibbs measures}.





 We are interested in the macroscopic behavior of the elongation,
 momentum and energy of the particles, at time $t$, as
 $N\rightarrow\infty$. Notice that $t$ is already the macroscopic
 time, since we have already multiplied the generator by $N$.
Taking advantage of the one-dimensionality of the system, we will use
\emph{lagrangian} coordinates, i.e. our space variables will be given
by the lattice coordinates $\{1/N, \dots, (N-1)/N,1\}$.
Also observe that at this time scale, the generator of the process is
given by $N\mathcal G_N^{\tau(t)}$.

 Consequently, we introduce the (time dependent)
 empirical measures representing the
 spatial distribution (on the interval $[0,1]$) of these quantities:
\begin{equation*}
\eta_\alpha^N(dx,t):=\frac 1N\sum_{i=1}^N\delta\left(x-\frac
  {i}{N}\right)\zeta_{i,\alpha}(t)\;dx,
\quad{\text{for}}\quad\alpha=1,2,3 . 
\end{equation*}


We expect the measures $\eta_\alpha^N(dx,t), \;\alpha=1,2,3$ to
converge, as $N\rightarrow\infty$, to measures \index{{\large{CHAPTER
      2:}}!$\mathfrak r(x,t)$}  $\mathfrak r(x,t)dx$,  \index{{\large{CHAPTER
      2:}}!$\mathfrak p(x,t)$}  \index{{\large{CHAPTER 2:}}!$\mathfrak
  r_0(x)$} \index{{\large{CHAPTER 2:}}!$\mathfrak p_0(x)$}  $\mathfrak
p(x,t)dx$, $-E(x,t)dx$ being  absolutely continuous with respect to
the Lebesgue measure and with density satisfying  the following system
of three conservation laws: 
  \begin{equation}\label{pde}
  \left\{\begin{array}{lcl}
  \partial_t \mathfrak r-\partial_x\mathfrak p=0&&\\
  \partial_t\mathfrak p-\partial_xP(\mathfrak r,\mathfrak e)=0&,&
  \left\{\begin{array}{l}
  \mathfrak r_0(x)=\mathfrak r(x,0),\;\mathfrak p_0(x)=\mathfrak p(x,0),\;E_0(x)=E(x,0)\\\\
  \mathfrak p(0,t)=0,\,P(\mathfrak r(1,t), \mathfrak e(1,t))=\tau(t)\\
  \end{array}
\right.\\
  \partial_t E-\partial_x(\mathfrak pP(\mathfrak r,\mathfrak e))=0&&
  \end{array}\right.
  \end{equation}
    for bounded, smooth initial data 
$\mathfrak r_0,\mathfrak p_0,E_0: [0,1]\rightarrow\mathbb R$ 
and the force $\tau(t)$ depending on time $t$.
Here we denoted by $\mathfrak r$ the specific volume, $\mathfrak p$ the
velocity, $E$ the total energy and $\mathfrak e:=E-\frac12 \mathfrak p^2$ the
internal energy.    

We need the solutions of the system \eqref{pde} to be
$C^2$-solutions. To assure this, the following additional
compatibility conditions at the  space-time edges $(x,t)=(0,0)$ and
$(x,t)=(1,0)$ have to be satisfied:
\begin{eqnarray}\label{compatibility1}
\lim_{x\rightarrow0}\mathfrak p_0(x)
=\mathfrak p(0,0)=0&,&\lim_{x\rightarrow1}P(\mathfrak r_0(x),\mathfrak e_0(x))=\tau(0)
\\
\label{compatibility2}
\lim_{x\rightarrow0}\frac d{dx}P(\mathfrak r_0(x),\mathfrak e_0(x))=0&,&\lim_{x\rightarrow1}\frac d{dt}P(\mathfrak r_0(x),\mathfrak e_0(x))=\tau^{\prime}(0)
\\
\label{compatibility3}
\lim_{x\rightarrow0}\frac {d^2}{(dt)^2}\mathfrak p_0(x)=0&,&\lim_{x\rightarrow1}\frac {d^2}{(dt)^2}P(\mathfrak r_0(x), \mathfrak e_0(x))=\tau^{\prime\prime}(0).
\end{eqnarray}

A proof of this can be adapted from Chapters 4.3, 7.5 and 3.5 of \cite{tatsien}.

For any test function $J:[0,1]\rightarrow\mathbb R$ with compact support  in $(0,1)$ consider the empirical densities 
 \begin{equation}
 \label{empdens1}\eta_\alpha^N(t,J):=\langle \eta_\alpha^N(dx,t);J\rangle\,=\,\frac 1N\sum_{i=1}^NJ\left(\frac{i}N\right)\zeta_{\alpha,i}(t).
  \end{equation}
Our  goal is to show  that,  starting with an initial distribution such that  there exist smooth functions $\mathfrak r_0$, $\mathfrak p_0$ and $E_0$  satisfying
 \begin{equation}\label{limit0}
 \{\eta_1^N(0,J),\eta_2^N(0,J),\eta_3^N(0,J)\} \rightarrow
 \left\{\int J(x)\mathfrak r_0(x)dx,\int J(x)\mathfrak p_0(x)dx, - \int
   J(x)E_0(x)dx\right\} 
 \end{equation} 
in probability as $N\rightarrow\infty$,  then at time $t\in[0,T]$ we  have the same convergence of $\eta^N_{\alpha}(t,J)$, $\alpha=1,2,3$  to the corresponding profiles $\mathfrak r(x,t)$, $\mathfrak p(x,t)$ and $E(x,t)$ respectively, 
that satisfy \eqref{pde}--\eqref{compatibility3}.

Here is the precise statement of our main result, where we make a
stronger assumption on the initial measure:

{ \theorem[Main Theorem]\label{conservation}
  For  any time  $t\in[0,T]$, denote by $\mu_t^N$ the probability
  measure on the path space $C([0,T],(\mathbb R^2)^N)$ of our process with
  generator $N\mathcal G_N$, 
  and starting from  the
  local  Gibbs measure $\nu^N_{\boldsymbol\lambda(\cdot,0)}$ 
  corresponding to the initial profiles $\tilde{\mathfrak u}_0$.
  Then for any smooth function $J:[0,1]\rightarrow\mathbb R$ and any
  $\delta>0$ 
   \begin{equation}
 \lim_{N\rightarrow\infty}\mu_t^N\left[\left|\frac 1N\sum_{i=1}^N
     J(\frac iN) \boldsymbol\zeta_{i}-\int_0^1J(x)\tilde{\mathfrak
     u}(x,t)dx\right|>\delta\right]=0. 
\end{equation}
where  $\mathfrak u$ is a $C^2$-solution to the system of conservation laws \eqref{pde}--\eqref{compatibility3} and $0<T<t_s$, $t_s$ being the time at which the solution $\mathfrak u$ produces the first shock.
}
\bigskip

{\remark\label{rem1} As our proof is based on the relative entropy method of \cite{yau}, it is only valid as long as  the solution to \eqref{pde} are $C^2$. Since, even for smooth initial data, the solution will develop shocks, we are forced to restrict our derivation to a time $0<T<t_s$, where $t_s$ is the time when the solution to the system of conservation laws enters the first shock.}

 {\remark\label{rem2}A  proof for the existence of smooth solutions to the initial-boundary-value problem \eqref{pde} can be found in  chapter 4.3, 7.5 and 3.5 of \cite{tatsien}. Notice that   we can rewrite the pressure $P$ as a function  of specific volume $\mathfrak r$ and entropy $\mathfrak s$:
 $$
\tilde P(\mathfrak r,\mathfrak s):=P(\mathfrak r,\mathfrak e).
 $$
Then we can rewrite  the initial boundary value problem  {\eqref{pde}}, in the smooth regime, in terms of the unknown $\mathfrak r$, $\mathfrak p$ and $\mathfrak s(\mathfrak r,\mathfrak e)$ as follows:
 \begin{equation}\label{pdes}
  \left\{\begin{array}{l}
  \partial_t \mathfrak r-\partial_x\mathfrak p=0\\\\
  \partial_t\mathfrak p-\partial_x\tilde P(\mathfrak r,\mathfrak s)=0\\\\
  \partial_t \mathfrak s=0
  \end{array}\right.
  ,  \left\{\begin{array}{l}
 \mathfrak r_0(x)=\mathfrak r(x,0),\;\mathfrak p_0(x)=\mathfrak p(x,0),\;\mathfrak s_0(x)=\mathfrak s(x,0)\\\\
  \mathfrak p(0,t)=0,\,\tilde P(\mathfrak r(1,t), \mathfrak s(1,t))=\tau(t)
  \end{array}\right.,
  \end{equation}
where we used the thermodynamic  relation 
$$
\tilde P(\mathfrak r,\mathfrak s)=-\frac {\partial \mathfrak e(\mathfrak r,\mathfrak s)}{\partial \mathfrak r}.
$$
Hence the specific entropy $\mathfrak s$ does not change in time and for any $x\in [0,1]$ is given through the initial data $\mathfrak s(x,0):=\mathfrak s_0(x)$.

In the non-consevative form, equation \eqref{pdes} reads as:
$$
\partial_t\begin{pmatrix}
\mathfrak r\\
\mathfrak p\\
\mathfrak s
\end{pmatrix}-\mathbf A(\mathfrak r,\mathfrak p,\mathfrak s)\partial_x\begin{pmatrix}
\mathfrak r\\
\mathfrak p\\
\mathfrak s
\end{pmatrix}=0
$$
where the $3\times3$-matrix $\mathbf A$ is defined by
$$
\mathbf A:=
\begin{pmatrix}
0&1&0\\
\frac{\partial{\tilde P}}{\partial \mathfrak r}&0&\frac{\partial\tilde{ P}}{\partial \mathfrak s}\\
0&0&0\
\end{pmatrix}=\mathbf S\cdot\begin{pmatrix}
c&0&0\\
0&-c&0\\
0&0&0
\end{pmatrix}\cdot\mathbf S^{-1}
$$
with $c:=c(\mathfrak r,\mathfrak s)=\sqrt{\frac{\partial{\tilde P}}{\partial \mathfrak r}}$ and
$$
\mathbf S:=\mathbf S(\mathfrak r,\mathfrak p,\mathfrak s)=\begin{pmatrix}
1&1&-\frac1c\frac{\partial{\tilde P}}{\partial \mathfrak s}\\
c&-c&0\\
0&0&c
\end{pmatrix}
.$$
 
 With these notations we can rewrite \eqref{pdes} in the characteristic form
 $$
 \mathbf S^{-1}\cdot\partial_t\begin{pmatrix}
\mathfrak r\\
\mathfrak p\\
\mathfrak s
\end{pmatrix}-\begin{pmatrix}
c&0&0\\
0&-c&0\\
0&0&0
\end{pmatrix}\cdot\mathbf S^{-1}\cdot\partial_x\begin{pmatrix}
\mathfrak r\\
\mathfrak p\\
\mathfrak s
\end{pmatrix}=0
 $$
 $$
 \Rightarrow
 \Bigg\{
 \begin{array}{l}
 c(\partial_t \mathfrak r-c\partial_x\mathfrak r)+(\partial_t \mathfrak p-c\partial_x\mathfrak p)+\frac 1c\frac{\partial\tilde P}{\partial s}(\partial_t \mathfrak s-c\partial_x\mathfrak s)=0\\
 c(\partial_t \mathfrak r+c\partial_x\mathfrak r)-(\partial_t \mathfrak p+c\partial_x\mathfrak p)+\frac1c\frac{\partial\tilde P}{\partial s}(\partial_t \mathfrak s+c\partial_x\mathfrak s)=0\\
\partial_t\mathfrak s=0.
 \end{array}
 $$
 In this way we  can apply the existence proof for $C^2$ solutions to \eqref{pde}--\eqref{compatibility3} for short times from  \cite{tatsien}.}

 \section{The Hydrodynamic Limit\label{HDL}}
 
 \subsection{The Relative Entropy}
 
 On the phase space $(\mathbb R^2)^N$ we now have two time--dependent families of probability measures. 
 One of them is the local Gibbs measure 
 $\nu^N_{\boldsymbol\lambda(\cdot, t)}$
 constructed from the solution of the system of conservation laws
 \eqref{pde}--\eqref{compatibility3}.  
 We denote its density by
 \begin{equation}
   \label{eq:gdensity}
   g^N_t = \prod_{i=1}^N 
   e^{\boldsymbol\lambda(i/N,t) \cdot \bzeta_i -\Theta(\boldsymbol\lambda(i/N,t))}
 \end{equation}

 On the other hand we have the actual distribution, whose density
  $f^N_t(\mathbf r,\mathbf p)$ is a solution, in the sense of distributions, of the Kolmogorov forward equation:
 \begin{equation}\label{kolmo}
 \left\{\begin{array}{ll}
  \frac{\partial  f_t^N}{\partial t}(\mathbf r,\mathbf p)&=N\mathcal
  G_N^{\tau(t), \star}f^N_t(\mathbf r,\mathbf p)\\\\
  f_0^N(\mathbf r,\mathbf p)&=g_0^N(\mathbf r,\mathbf p).
 \end{array}\right.
 \end{equation}
 By $\mathcal G_N^{\tau,\ast}=L_N^{\tau,\star}+\gamma S_N$ we denote the
 adjoint operator of $ \mathcal G_N$ with respect to the Lebesgue measure, where $L_N^{\tau,\star}$   can be computed as
 $L^{\tau,\star}_N=-L^{\tau}_N$.    

The relative entropy of $f^N_t$ with respect to $g_{t}^N$
is defined by \index{{\large{CHAPTER 2:}}!$H_N(t)$}
\begin{equation}
  \label{eq:relentropy}
  H_N(t) = \int  f^N_t \;\log{\frac{f^N_t}{g^N_{t}}}\;
  d\mathbf r d\mathbf p
\end{equation}

Our main result will follow from:

{\theorem[Relative entropy]\label{relentr}Under the same assumptions as  in Theorem \ref{conservation}, for any  time $t\in[0,T]$  \index{{\large{CHAPTER 2:}}!$H_N\left(\nu_t^N| \nu^N_{\mathfrak u(\cdot,t)}\right)$}
$$
\lim_{N\rightarrow\infty}\frac 1N H_N\left(t\right)
=0.
$$
}
  To see how Theorem \ref{relentr} implies the Main Theorem, see
  \cite{kipnis, stefano}

\textbf{Remark:} 
Recall that the relative entropy $H(\alpha|\beta)$  of a probability measure $\alpha$ with respect to a probability measure $\beta$ can be rewritten as
\begin{equation}
H(\alpha|\beta)=\sup_\varphi\left\{\int\varphi d\alpha-\log\int
  e^{\varphi}d\beta\right\} 
\end{equation}
where the supremum is taken over all bounded measurable functions $\varphi$.
It is easy to see that the relative entropy  has the following
properties: $H(\alpha|\beta)$ is positive convex and lower semicontinuous. 
It follows that for any measurable function $F$ and any $\sigma>0$:
 \begin{equation}\label{reineq}
 \int F\; d\alpha\leq\frac1{\sigma}\log \int e^{\sigma F}\; d\beta + 
\frac 1 \sigma H(\alpha|\beta).
 \end{equation}

\subsection{Time Evolution of the Relative Entropy\label{RE}}

In this Section we will prove Theorem \ref{relentr}.

Notice that with the choice of our initial distribution
 $$H_N(0)=0.$$
The strategy is to show that for some constant $C$ 
\begin{equation}
H_N(t)\leq C\int_0^tH_N(s)ds+\int_0^tR_N(s)ds
\end{equation}
with
\begin{equation}\label{RN}
\lim_{N\rightarrow\infty}\frac1N\int_0^{t}R_N(s)ds=0.
\end{equation}
Then it follows by  Gronwall's inequality that $\lim_{N\rightarrow\infty}\frac{H_N(t)}{N}=0$ which concludes the proof of Theorem \ref{relentr}.
We first prove the  following inequality:

{\begin{lemma}\label{dineq}
\begin{equation}
\label{inequality}
H_N(t)\leq-\int_0^t ds \int f_s^N \left(N\mathcal
    G^{\tau(s)}_N + \partial_s \right)\log g^N_s 
  d\mathbf r\;d\mathbf p
\end{equation}
 
\end{lemma}}
\begin{proof}
 Observe that we can rewrite the time increment of the entropy as
\begin{eqnarray*}
 H_N(s+h)-H_N(s)&=& \int f_{s+h}^N \log f_{s+h}^N d\mathbf r\; d\mathbf
   p - \int f_{s}^N \log f_{s}^N d\mathbf
   r\;d\mathbf p\\ 
&&-\int f_{s+h}^N\log g_{s+h}^N d\mathbf r\; d\mathbf p 
+\int f_{s}^N \log g_{s}^N d\mathbf r\; d\mathbf p
\end{eqnarray*}
By convexity of the function $\phi(f)=f\log f$, since Lebesgue measure
is stationary for the dynamics generated by $\mathcal G_N^{\tau(t)}$, we have that 
\begin{eqnarray}\label{entdec}
\int f_{s+h}^N \log f_{s+h}^N d\mathbf r\; d\mathbf
   p \le \int f_{s}^N \log f_{s}^N d\mathbf
   r\;d\mathbf p
\end{eqnarray}
Denote by $T_{s,s+v}^N$  the evolution, from time $s$ to some time
$s+v$ corresponding to the generator $N\mathcal G_N^{\tau(h)}$. Then
the probability density evolves as
\begin{equation}\label{evolutionT}
T_{s,s+v}^{N,\star}f_s^N=f_{s+v}^N.
\end{equation}
where $T_{s,s+v}^{N,\star}$ is the adjoint of $T_{s,s+v}^{N}$ with
respect to the Lebesgue measure. 
Then, since $g_s$ is smooth, we have
$$
T_{s,s+h}^N\log g_{s+h}^N- \log g_s^N =\int_0^h
T_{s,s+v}\left[\left(NG_N^{\tau(v)}+\frac{\partial}{\partial v}\right) 
\log g_{s+v}^N\right]dv. 
$$
Hence with \eqref{entdec}:
\begin{multline*}
H_N(s+h)-H_N(s)\leq -\int f_{s}^N \left(T_{s,s+h}^N\log g_{s+h}^N-\log g_s\right)
d\mathbf r\,d\mathbf p\\
= -\int_0^h dv \int f_{s+v}^N \left(N\mathcal
  G_N^{\tau(v)}+\frac{\partial}{\partial v}\right) \log g_{s+v}^N d
\mathbf r\,d\mathbf p \ .
\end{multline*}

We conclude the proof by setting $s=0$ since $H^N(0)=0$.
\end{proof}

\bigskip

Before we proceed in the proof, we have to introduce some further notations. 
 For any $C^1$ function $F:=(f_1,f_2,f_3)^T:\mathbb R^3\rightarrow\mathbb R^3$ we define
\begin{equation*}
DF(\mathbf a):=\left((Df_1)(\mathbf a),(Df_2)(\mathbf a),(Df_3)(\mathbf a)\right)^T,
 \end{equation*}
with $Df_i(\mathbf a)$, $i=1,2,3$ defined by \eqref{D}.
Recall that $\tilde{\mathfrak u} = (\mathfrak r, \mathfrak p, -E)^T$,
 $\mathfrak e: = E-\frac{1}{2} \mathfrak p^2$ and let us denote by
\begin{equation}\label{macroflux}
 \tilde{\mathbf J}(\tilde{\mathfrak u}):= (\mathfrak p,P(\mathfrak r,\mathfrak e),
 -\mathfrak pP(\mathfrak r,\mathfrak e))^T = \left(\mathfrak p, \frac{\lambda_1(\mathfrak
     r,\mathfrak e)}{\lambda_3(\mathfrak r,\mathfrak e)}, -\mathfrak p\frac{\lambda_1(\mathfrak r,\mathfrak e)}{\lambda_3(\mathfrak r,\mathfrak e)}\right)^T
 \end{equation} 
 the flux of  \eqref{pde}, that can be rewritten as
 $$
 \partial_t\tilde{\mathfrak u} =D\tilde{\mathbf J}(\tilde{\mathfrak
   u}) \partial_x\tilde{\mathfrak u} 
 $$
  with  the  Jacobian 
  \begin{equation}\label{macrojacflux}
  D\tilde{\mathbf J}(\tilde{\mathfrak u})\begin{pmatrix}
  0&1&0\\\\
  \frac {\partial P}{\partial \mathfrak r}&-\mathfrak p\frac {\partial P}{\partial \mathfrak e }&\frac {\partial P}{\partial \mathfrak e}\\\\
  -\mathfrak p\frac {\partial P}{\partial \mathfrak r}& 
  - P + \mathfrak p^2\frac {\partial P}{\partial \mathfrak e }
  &-\mathfrak p\frac {\partial P}{\partial  \mathfrak e}
  \end{pmatrix}
  \end{equation}
 With the dual relation \eqref{dual}, $\boldsymbol\lambda$ is solution of the symmetric system
 \begin{equation}\label{sympde}
 \partial_t[D\Theta(\boldsymbol\lambda)]=\partial_x[D\Sigma(\boldsymbol\lambda)],
 \end{equation}
 where 
 $$
 \Sigma(\boldsymbol\lambda)=\boldsymbol\lambda\cdot\tilde{\mathbf J}(D\Theta(\boldsymbol\lambda)).
 $$
 Equation \eqref{sympde} can be rewritten as
 $$
 (D^2\Theta)\partial_t\boldsymbol\lambda=(D^2\Sigma)\partial_x\boldsymbol\lambda.
 $$
 Since 
 $$
 D^2\Theta(\boldsymbol\lambda(t,x))^{-1}=(D^2\Phi)(\tilde{\mathfrak u}(t,x)),
 $$
 it follows that
 $$
 \partial_t\boldsymbol\lambda(D^2\Phi)=(D^2\Sigma)\partial_x\boldsymbol \lambda.
 $$
 Since
 $$
 (D^2\Sigma)=(D^2\Phi)(D\tilde{\mathbf J}(\tilde{\mathfrak u}))
 $$
 the following system of partial differential equations is satisfied:
 \begin{equation}\label{dualpde}
\partial_t\boldsymbol \lambda(t,x)=(D\tilde {\mathbf J})^T(\tilde {\mathfrak
  u})\partial_x\boldsymbol\lambda(t,x) .
\end{equation}

\bigskip


Let us define the microscopic fluxes:
\begin{equation}
  \label{eq:micflux}
  \begin{split}
    \mathbf J_{i-1,i} &:=(- p_{i-1}, - V^{\prime}(r_i), p_{i-1}V^{\prime}(r_i))^T
    \qquad i=1,\dots, N-1,\\
     \mathbf J_{N,N+1} &:= (-p_{N}, -\tau(t) , p_{N}\tau(t))^T
  \end{split}
\end{equation}
By the definition of the Liouville operator given by \eqref{liouville}, 
$$
L_N^{\tau(t)} \bzeta_i =  \mathbf J_{i-1,i} -  \mathbf J_{i,i+1}.
$$
Finally let us define
$$
\mathbf v_j:=(0,p_j,-p_j^2/2)^T.
$$
Hence with the definition of the symmetric operator given by \eqref{sym},
\begin{eqnarray*}
S_N(\boldsymbol \zeta_j)=-2\mathbf v_j+\mathbf v_{j+1}+\mathbf v_{j-1},\quad j=2,\dots,N-1\\
S_N(\boldsymbol \zeta_N)=-\mathbf v_N+\mathbf v_{N-1},\quad S_N(\boldsymbol \zeta_1)=-\mathbf v_1+\mathbf v_{2}\\
\end{eqnarray*}
{\begin{lemma}\label{liouvilleterm1}
\begin{equation}
 NL_N^{\tau(t)}\log g^N_t
 = \sum_{i=1}^{N} \partial_x\boldsymbol \lambda(\frac iN,t)
 \cdot\mathbf J_{i-1,i} + N\lambda_2(1,t)\tau(t) +a_N(t)
\label{eq:basic}
\end{equation}
where $a_N(t)$ is such that 
 $$
\lim_{N\rightarrow{\infty}}\frac1N\int_0^{t}\int
 a_N(s) f^N_s d\mathbf p d\mathbf r  ds =0
$$
\end{lemma}}

 \begin{proof}
  \begin{eqnarray*}
&& N L_N^\tau\log g^N_t (\mathbf r, \mathbf p) = 
N \sum_{i=1}^N \blambda (\frac iN,t)\cdot 
\left(\mathbf J_{i-1,i} -  \mathbf J_{i,i+1}\right) \\\\  
&=& N \sum_{i=1}^N \left(\blambda (\frac iN,t) -
  \blambda(\frac{i-1}N,t)\right) \cdot \mathbf J_{i-1,i} 
-\blambda(0,t)\cdot \mathbf J_{0,1} + \blambda(1,t)\cdot \mathbf J_{N.N+1}
 \end{eqnarray*}
Taking into account the boundary conditions on $\blambda$ we have
\begin{equation}
  \label{eq:bleft}
  \blambda(0,t)\cdot \mathbf J_{0,1} = \lambda_2(0,t) V^{\prime}(r_1) =
  \mathfrak p(0,t) V^{\prime}(r_1) = 0
\end{equation}
and 
\begin{equation}
  \label{eq:bright}
  \blambda(1,t)\cdot \mathbf J_{N.N+1} =  - p_N\lambda_1(1,t)
  - \lambda_2(1,t)\tau(t) + \lambda_3(1,t)\tau(t)p_N = 
   - \lambda_2(1,t)\tau(t) 
\end{equation}
because $\tau(t)\lambda_3(1,t)=\lambda_1(1,t)$.
 Since $\blambda$ is a $C^2$-functions, we obtain (\ref{eq:basic}) 
  with
$$
\left| a_N(t)\right| = \frac C{N} \sum_{i=1}^{N-1}\|\mathbf J_i\|
 $$
 It remains to show, that
 $\lim_{N\rightarrow\infty}\int_0^t\int\frac{a_N(s)}{N}d\nu_s^Nds=0$. This
 will be an easy consequence of  Lemma \ref{energybound}.
 \end{proof}
 
 {\lemma\label{liouvilleterm2}
 $$
 \partial_t\log g^N_{t}= 
\sum_{i=1}^N(D\tilde {\mathbf J})^T \left(\tilde {\mathfrak u}(\frac
 iN,t)\right) \partial_x  \boldsymbol\lambda (\frac iN,t) 
\cdot  \left({\boldsymbol\zeta}_i-\tilde{\mathfrak u}(\frac iN,t)\right) 
 $$
 \begin{proof}
 \begin{eqnarray*}
 \frac{\partial}{\partial t}\log g^N_{ t}&=&
 \frac{\partial}{\partial t}
 \sum_{i=1}^N\left(\boldsymbol\lambda(\frac iN,t)
\cdot{\boldsymbol\zeta}_i- \Theta\left(\boldsymbol\lambda(\frac iN,t)\right)\right)\\ 
 &=&\sum_{i=1}^N
\partial_t\boldsymbol\lambda (\frac iN,t)\cdot
\left({\boldsymbol\zeta}_i- 
D\Theta\left(\boldsymbol\lambda(\frac iN,t)\right)\right)\\
  \end{eqnarray*}
By (\ref{dual}), $D\Theta\left(\boldsymbol\lambda(\frac iN,t)\right)= \tilde{\mathfrak u}(\frac iN,t)$,
and \eqref{dualpde} the result follows.
 \end{proof}
 
{\lemma\label{noise} Recall the definition of the symmetric operator
  given by \eqref{sym}.
}
$$
- \lim_{N\rightarrow\infty} \frac1N \int_0^{t}\int \left(N S_N\log
 g^N_s\right) f^N_s  d\mathbf p d\mathbf r\;   ds
 \leq \lim_{N\rightarrow\infty}\frac1{\sigma N}\int_0^tH_N(s)ds,
  $$
  where $\sigma$ is a constant  independent of $N$ with
  $0<\sigma<\sup_{x, s} \frac 1{2\lambda_3(x,s)}$. 
}

 \begin{proof}
 
\begin{eqnarray*}
&&\!\!\!\!\!\!\!\!S_N\log g_s^N\\
 &&\!\!\!\!\!\!\!\!=\sum_{i=2}^{N-1}\blambda(\frac iN,t)\cdot(\mathbf v_{i-1}-2\mathbf
  v_i+\mathbf v_{i+1}) + \blambda(\frac1N,t)\cdot(-\mathbf v_1+\mathbf v_2) +
  \blambda(1,t)\cdot(\mathbf v_{N-1}-\mathbf v_N)\\
&&\!\!\!\!\!\!\!\!=\sum _{i=2}^{N-1} \left(\blambda(\frac {i-1}N,t)-2\blambda(\frac
  iN,t)+\blambda(\frac {i+1}N,t)\right)\cdot \mathbf v_i\\ 
&&\qquad\quad\qquad\qquad\qquad+\left(\blambda(\frac2N,t)
  -\blambda(\frac1N,t)\right)\cdot \mathbf v_1
+\left(\blambda(\frac{N-1}N,t)-\blambda(1,t)\right)\cdot \mathbf v_N 
 \end{eqnarray*}
 
In Lemma \ref{energybound} we will show that the expectation  of  
$\frac 1N\sum_i\|\mathbf v_i\|$ is uniformly bounded for all  $N$ and
hence, since $\blambda$ is in $C^2$, the first term vanishes in the
limit as $N\rightarrow\infty$.  

Recall that by the entropy inequality \eqref{reineq}, for any $\sigma>0$ we have for $k\in\{1,\dots,N\}$: 
\begin{eqnarray*}
\frac1N \int p_k^2  f^N_s  d\mathbf p d\mathbf r  \leq
\frac1{N\sigma}\log\int e^{\sigma p^2_k} \nu^N_{\boldsymbol\lambda(\cdot,s)} 
+\frac 1{N\sigma}H(s)
\end{eqnarray*}
Since this inequality is true for any $\sigma>0$, the integral on the
right hand side of the inequality is bounded as long as
$\sigma<\sup_x \frac 1{2\lambda_3(x,s)}$ and hence the first  term vanishes as
$N\rightarrow\infty$. The expected value of $p_k$ can be controlled in
a similar way. 
\end{proof}
 
\rm
 

So far we have from Lemma \ref{liouvilleterm1}, \ref{liouvilleterm2} and \ref{noise} 
  \begin{eqnarray}\label{1}
\nonumber H_N(t)\!\!\!\!
 \nonumber &\leq&\!\!\!\int_0^t\int\sum_{i=1}^N \partial_x
   \boldsymbol\lambda (\frac iN,s) \left[\mathbf
   J_{i-1,i}-(D\tilde{\mathbf J})^T(\tilde{\mathfrak u}(\frac
   iN,s))\left({\boldsymbol\zeta}_i-\tilde{\mathfrak u}(\frac
     iN,s)\right)\right]f^N_s  d\mathbf p\; d\mathbf r\;ds\\ 
  &&-\int_0^t  N \tau(s) \lambda_2(1,s)ds + \frac1{\sigma}\int_0^tH_N(s)ds+\int_0^tR_N(s)ds
  \end{eqnarray}
  where $R_N(t)$ is such that
  \eqref{RN} holds. 
  
  By \eqref{macroflux} we have 
  $$
  \int_0^1\partial_x \boldsymbol\lambda(x,t)\cdot\tilde{ \mathbf J}(\tilde{\mathfrak u}(x,t))dx=\int \frac{\partial}{\partial x}\left(\frac{\lambda_1(x,t)\lambda_2(x,t)}{\lambda_3(x,t)}\right)dx=\tau(t)\lambda_2(1,t). 
  $$
 and consequently we can replace $- N \tau(t) \lambda_2(1,t)$ by 
  $$
  - \sum_{i=1}^N \partial_x \boldsymbol\lambda(\frac iN,t) \cdot 
  \tilde{\mathbf J}(\tilde{\mathfrak u}(\frac iN,t))
  $$
  with an error uniformly bounded in $N$.
    It follows that  from \eqref 1  we have 
  \begin{multline}\label{2}
 H_N(t)\leq
  \int_0^t\int\sum_{i=1}^N \partial_x \boldsymbol\lambda (\frac iN,s)\times\\ 
\left[\mathbf J_{i-1,i}-\tilde{\mathbf J}\left(\tilde{\mathfrak u}(\frac
    iN,s)\right) -(D\tilde{\mathbf J})^T(\tilde{\mathfrak
    u}(\frac iN,s))\left({\boldsymbol{\zeta}}_i-\tilde{\mathfrak u}(\frac
    iN,s)\right)\right] f^N_s  d\mathbf p\; d\mathbf r\;ds \\ 
+\frac1{\sigma}\int_0^tH_N(s)ds + \int_0^t R_N(s)ds.
  \end{multline}
  
Our next goal is to prove a weak form of local equilibrium. In view of
this we introduce microscopic averages over blocks of size $2k+1$: In
what follows, for any  vector field $\mathbf
Y_i:=(Y_{1,i},Y_{2,i},Y_{3,i})^T: (\mathbb R^2)^3\rightarrow\mathbb
R^3$ we denote by $\mathbf Y_i^k:=(Y^k_{1,i},Y^k_{2,i},Y^k_{3,i})^T$,
block averages over blocks of length $2k+1$, where $k>0$ is
independent of $N$. For example 

\index{{\large{CHAPTER 2:}}!$\boldsymbol\zeta_i^k$}
\begin{equation}\label{empdens}
\boldsymbol\zeta_i^k=(\zeta_{1,i}^k,\zeta_{2,i}^k,\zeta_{3,i}^k)^T:=(r_{i}^k,p_{i}^k,-e_i^k)^T:=\frac 1{2k+1}\sum_{|i-l|\leq k}\boldsymbol\zeta_l. 
\end{equation}
These blocks are microscopically large but on the macroscopic scale
they are small, thus $N$ goes to infinity first and then $k$ goes to
infinity. We also need to introduce another small parameter $\ell$ and
consider small macroscopic blocs of length $\ell N$ at the boundaries.

 For any smooth  and bounded function $F:[0,1]\rightarrow\mathbb R$
 and any bounded  function $\psi:\mathbb R^2\rightarrow\mathbb R$, we obtain the following summation by parts formula
\begin{equation}\label{sbp4}
\frac 1N\sum_{i=1}^NF(\frac iN)\psi(r_i,p_i)=\frac 1N\sum_{i=[N\ell]}^{N-[N\ell]}F(\frac iN)\frac1{2k+1}\sum_{|j-i|\leq k}\psi(r_j,p_j)+\mathcal O(\frac {k+N\ell}N).
\end{equation}
 Here  we first restricted the sum to configurations over $\{[N\ell],\dots,N-[N\ell]\}$, for some small  $\ell>0$, such that $\ell\rightarrow0$ after $N\rightarrow\infty$ and $\ell N>>k $. In this way, we avoid touching the boundary when we introduce the block averages. The error we made   will vanish in the limit since $\ell\rightarrow 0$. 
\medskip

We also need to do some cut off in order to have only bounded variables: \\
Let $\mathcal C_{i,b}:=\{e_{i-1},e_i\leq b\}$, and define 
$$
\mathbf J_{i-1,i}^b:=\mathbf J_{i-1,i}\mathbf 1_{\mathcal C_{i,b}}\qquad{\text{and}}\qquad{\boldsymbol\zeta}_{i}^b:={\boldsymbol \zeta}_i\mathbf1_{\mathcal C_{i,b}},
$$ 
then these functions are bounded. Also denote 
$ \mathbf{\tilde J}^b(\tilde{\mathfrak u})$ 
the corresponding expectation with respect to the Gibbs measure of
parameters $\boldsymbol\lambda(\tilde{\mathfrak u})$, that converges
to $ \mathbf{\tilde J}(\tilde{\mathfrak u})$ as $b\to \infty$.

Assumptions \eqref{V1} 
on the potential assert that by
the entropy inequality \eqref{reineq} with reference measure
$d\nu^N_{\blambda(\cdot,t)}$,  the error we make by the replacement of
$\mathbf J_{i-1,i}$ and ${\boldsymbol \zeta}_i$ by $\mathbf
J_{i-1,i}^b$ and ${\boldsymbol\zeta}_{i}^b$ respectively is small in
$N$ if we can show that $\frac 1NH_N(s)\rightarrow 0$  as
$N\rightarrow$ 0:\\ For any $\sigma>0$ small enough 
\begin{eqnarray}
&&\int \sum_{i=1}^{N}\partial_x\boldsymbol\lambda(\frac iN,s)\mathbf J_{i-1,i}\mathbf 1_{\mathcal C_{i,b}^c}d\nu_s^N\nonumber\\
&\leq&\frac 1{\sigma}\sum_{i=1}^N\log\left(\int
  e^{\sigma\partial_x\boldsymbol\lambda(\frac
    iN,s)\mathbf J_{i-1,i}\mathbf 1_{\mathcal
      C_{i,b}^c}} d\nu_{\blambda(\cdot,t)}\right)+\frac {H_N(s)}{\sigma}\nonumber\\ 
&\leq&\frac 1{\sigma}\sum_{i=1}^N\log\left(1+\int_{\mathcal C_{i,b}^c}
  e^{\sigma\partial_x\boldsymbol\lambda(\frac
    iN,s)\mathbf J_{i-1,i}} d\nu_{\blambda(\cdot,t)}\right)+\frac {H_N(s)}{\sigma}\nonumber\\ 
&=&\frac {NC(b,\sigma)}{\sigma}+\frac {H_N(s)}{\sigma}\label{12}
\end{eqnarray}
where $\lim_{b\rightarrow\infty}C(b,\sigma)=0$ for any $\sigma>0$.

Using that  $\boldsymbol\lambda$ and $\mathfrak u$ are in $C^2$  and
formula \eqref{sbp4}, we arrive at 

\begin{eqnarray*}
&&\sum_{i=1}^N\partial_x \boldsymbol\lambda(\frac iN,s)\left[\mathbf
  J_{i-1,i}^b-\tilde{\mathbf J}^b \left({\tilde{\mathfrak u}(\frac
      iN,s)}\right)-(D\tilde{\mathbf J})^T \left(\tilde{\mathfrak
      u}(\frac iN,s)\right)\left({\boldsymbol\zeta}_{i}^b
    -\tilde{\mathfrak u}(\frac iN,s)\right)\right]\\ 
&=&\sum_{[N\ell]}^{N-[N\ell]}\partial_x \boldsymbol\lambda(\frac {i}N,s)\times\\
&&\Big[\frac1{ 2k+1}\sum_{|l-i|\leq  k}\mathbf J_{l-1,l}^b -
\tilde{\mathbf J}^b \left(\tilde{\mathfrak u}(\frac {i}N,s)\right) -
(D\tilde{\mathbf J})^T \left(\tilde{\mathfrak u}(\frac {i}N,s)\right)
\left({\boldsymbol\zeta}_{i}^{b,k}-\tilde{\mathfrak u}
  (\frac {i}N,s)\right) \Big]\\
&&+\mathcal O(k + N\ell)
\end{eqnarray*}


The following theorem will be proved
in Section \ref{OB}. 

{\theorem[The one-block estimate]\label{obe}
For any $\ell, b$: 
\begin{multline}\label{obe0}
\lim_{k\rightarrow \infty}\lim_{N\rightarrow\infty}
\frac1N\sum_{i=[N\ell]}^{N-[N\ell]} \int_0^t \int
\left|\frac1{ 2k+1}\sum_{|l-i|\leq
    k}\mathbf J^b_{l-1,l}- \tilde{\mathbf J}^b 
\left({\boldsymbol\zeta}_{i}^k\right)\right| f^N_s 
 d\mathbf p \; d\mathbf r \; ds = 0
\end{multline}
}
\medskip

With this Theorem we obtain:
\begin{multline}\label{4}
\frac{H_N(t)}N\leq
\frac 1N\sum_{i=[N\ell]}^{N-[N\ell]}\int_0^t
\int\boldsymbol\Omega\Big(\boldsymbol\zeta_{i}^{k},\tilde{\mathfrak u}(\frac{i}N,s)\Big) f^N_s 
 d\mathbf p \; d\mathbf r \; ds\\
+\int_0^t\frac{R_{N,k,\ell,b}(s)}Nds+\int_0^t\frac {H_N(s)}{N\sigma}  ds
\end{multline}
for some $\sigma>0$. $R_{N,k,\ell,b}$ is such that
$$
\lim_{b\rightarrow\infty}\lim_{\ell\rightarrow0}\lim_{k\rightarrow \infty}\lim_{N\rightarrow\infty}\int_0^t\frac{R_{N,k,\ell,b}(s)}{N}ds=0,
$$
and we used \eqref{dualpde} to define

$$
\boldsymbol\Omega(\mathfrak z,\tilde{\mathfrak u}):=\partial_x \boldsymbol\lambda\cdot\left(\tilde{\boldsymbol{\mathbf J}}\left(\mathfrak z\right)-\tilde{\mathbf J}\left(\tilde{\mathfrak u}\right)\right)-\partial_t \boldsymbol\lambda\cdot\left(\mathfrak z-\tilde{\mathfrak u}\right).
$$
Hence
\begin{equation}\label{115}
D_{\mathfrak z}\boldsymbol\Omega(\mathfrak z,\tilde{\mathfrak u})=\left((D\tilde{\mathbf J})^T(\mathfrak z)\cdot\partial_x \boldsymbol\lambda-\partial_t \boldsymbol\lambda\right)
\end{equation}
is equal to zero if $\mathfrak z$ is a solution of \eqref{dualpde} and consequently:
$$\boldsymbol\Omega(\tilde{\mathfrak u},\tilde{\mathfrak u})=0,\qquad D_{\mathfrak z}\boldsymbol\Omega(\tilde{\mathfrak u},\tilde{\mathfrak u})=0.
$$
Applying the entropy inequality  \eqref{reineq} on the sum in \eqref{4}, we obtain that for some $\sigma>0$ it is bounded above by
\begin{multline}\label{5}
\frac{1}{N\sigma}\int_0^t\log\int \exp
\left\{\sigma\sum_{i=[N\ell]}^{N-[N\ell]}
  \boldsymbol\Omega(\boldsymbol\zeta_{i}^k,\mathfrak u(\frac{i}N,s)) \right\}
 g_s^N d\mathbf p \; d\mathbf r \; ds + \frac1{N\sigma}\int_0^t H_{N}(s)
 ds
\end{multline}

 Hence it remains to prove, that the first term of this expression is of order $\mathcal O(\frac1N)$. This will be done using  the following special case of Varadhan's Lemma:

{\theorem[Varadhan's Lemma]\label{varadhan}  Let $\nu^n_{\boldsymbol \lambda}$ be the product homogenuous measure with marginals $\nu_{\boldsymbol\lambda}$ given by \eqref{eq:marginals} and with rate function $ I:\mathbb R^2\times\mathbb R_-\rightarrow\mathbb R$ defined by
$$
 I(\mathbf x):=\Phi(\mathbf x)-\mathbf x\cdot\boldsymbol\lambda+\Theta(\boldsymbol \lambda).
$$
Then for any bounded continuous function $F$ on $\mathbb R^2\times\mathbb R_-$ 
$$
\lim_{n\rightarrow\infty}\frac1n\log\int e^{nF(\boldsymbol\zeta)}d\nu^n_{\boldsymbol\lambda}=\sup_{\mathbf x}\{F(\mathbf x)-I(\mathbf x)\}
$$
}

\begin{proof}
A proof of this Theorem can be adapted from \cite{stefano, kipnis, varadhan}

\end{proof}

In order to apply this Theorem  we arrange  the sum in \eqref{5} as sums  over disjoint blocks and then take advantage of the fact that the local Gibbs measures are product measures:

Assume without loss of generality that $ 2k+1$ devides $N-2[N\ell]$, then
$$
\sum_{i=[N\ell]}^{N-[N\ell]}
\boldsymbol\Omega(\boldsymbol\zeta_{i}^k,\tilde{\mathfrak
  u}(\frac{i}N,s)) = 
\sum_{j\in\{- k,\dots, k\}}\sum_{i\in B_{[N\ell]}^{N-2[N\ell],k}}
\tau_j \boldsymbol
\Omega(\boldsymbol\zeta_{i}^k,\tilde{\mathfrak u}(\frac{i}N,s)) 
$$
where $B_{[N\ell]}^{N-2[N\ell],k}:=\left\{r( 2k+1)+[N\ell]+k;
  r\in\left\{0,\dots,\frac{N-2[N\ell]-2k}{ 2k+1}\right\}\right\}$.  In
this way, for any fixed $j$,  the terms in the sum over $i\in
B_{[N\ell]}^{N-2[N\ell],k}$ depend on configurations in  disjoint
blocks. Thus the random variables 
$$\tau_j\boldsymbol\Omega
(\boldsymbol\zeta_i^k,\tilde{\mathfrak u}(\frac{i}N,s))$$ 
 are independent under $\nu^N_{\boldsymbol\lambda(\cdot, s)}$.

Using H{\"o}lder inequality, the first term in \eqref{5} is bounded above by
\begin{multline*}
\frac{1}{N\sigma}\int_0^t\log\int \prod_{j\in\{- k,\dots, k\}}\exp \left\{\sigma \sum_{i\in B_{[N\ell]}^{N-2[N\ell],k}}\tau_j\boldsymbol\Omega  (\boldsymbol\zeta_i^k,\mathfrak u(\frac{i}N,s))\right\}
g_s^N d\mathbf p \; d\mathbf r \; ds\\
\leq\frac{1}{N\sigma( 2k+1)} \int_0^t\qquad\qquad\qquad\qquad\qquad\qquad\qquad\qquad\qquad\qquad\qquad\qquad\qquad\quad\qquad\\
\sum_{j\in\{- k,\dots, k\}}\log\int\exp \left\{\sigma( 2k+1) \sum_{i\in B_{[N\ell]}^{N-2[N\ell],k}}\tau_j\boldsymbol\Omega  (\boldsymbol\zeta_i^k,\tilde{\mathfrak u}(\frac{i}N,s))\right\}
g_s^N d\mathbf p \; d\mathbf r \; ds\\
=
\frac{1}{N\sigma( 2k+1)}\sum_{i=[N\ell]}^{N-[N\ell]}\int_0^t\log\int
\exp\left\{\sigma ( 2k+1)\boldsymbol\Omega
  (\boldsymbol\zeta_i^k,\tilde{\mathfrak u}(\frac{i}N,s))\right\} g_s^N d\mathbf p \; d\mathbf r \; ds.
\end{multline*}

Then, since all the functions in this expression are smooth and the family of local Gibbs measures converges weakly, we obtain that
\begin{multline*}
\!\!\!\!\!\!\!\lim_{k\rightarrow\infty}\!\lim_{N\rightarrow\infty}\frac{1}{(
  2k+1)N\sigma}\!\sum_{i=[N\ell]}^{N-[N\ell]}\!\int_0^t\log\int
\exp\left\{ \sigma( 2k+1) \boldsymbol\Omega
  (\boldsymbol\zeta_i^k,\tilde{\mathfrak u}(\frac{i}N,s)) \right\}g_s^N d\mathbf p\,d\mathbf r\,ds\\
=\lim_{k\rightarrow\infty}\frac1{\sigma(
  2k+1)}\int_0^t\int_0^1\log\int \exp\left\{ ( 2k+1)\sigma
  \boldsymbol\Omega  (\boldsymbol\zeta_i^k,\tilde{\mathfrak u}(x,s)) 
 \right\}d\nu_{\boldsymbol\lambda(x,s)}dxds.\\
\end{multline*}
So now  for each $x\in[0,1]$, the distribution of the particles in a box of size $k$ is given by the invariant Gibbs measure with average $\mathfrak u(x,s)$. such that we can apply Theorem \ref{varadhan} on this product measure to obtain that the last expression is equal to
\begin{equation}\label{6}
\frac1{\sigma}\int_0^t\int_0^1\sup_{\mathfrak z}\{\sigma\boldsymbol\Omega  \left(\mathfrak z,\tilde{\mathfrak u}(x,s)\right)-I(\mathfrak z)\}dx.
\end{equation}
To conclude Theorem \ref{relentr} it thus remains to show that this is
equal to zero.  
Since $I$ and $\boldsymbol\Omega  $ are both convex, and both
functions and their derivatives are vanishing at $\mathfrak
z=\tilde{\mathfrak u}$, it follows from   assumption \eqref{V1}
on the potential that  $\sigma\boldsymbol\Omega(\mathfrak
z,\tilde{\mathfrak u})\leq I(\mathfrak z)$ for $\sigma$ small
enough. Hence there exists a $\sigma$ such that the last expression is
equal to zero. 

This concludes the proof of Theorem \ref{relentr}:

Since
$$
H_N(t)\leq C\int_0^tH_N(s)ds +\int_0^tR_{N,k,\ell, b}(s)ds, 
$$
for some uniform constant $C$, it follows by Gronwall inequality that
\begin{eqnarray*}
H_N(t)&\leq& H_N(0)e^{Ct}+\int_0^tR_{N,k,\ell, b}(s)e^{C(t-s)}ds\\
&\leq&e^{Ct}\left(H_N(0)+\int_0^tR_{N,k,\ell, b}(s)ds\right).
\end{eqnarray*}
Hence the claim follows, since
$$
\lim_{b\rightarrow\infty} \lim_{\ell\rightarrow0}  
\lim_{k\rightarrow\infty} \lim_{N\rightarrow\infty}
\int_0^t\frac{R_{N,k,\ell, b}(s)}{N}ds=0. 
$$

\subsection{The one block estimate\label{OB}{ \rm{(Theorem \ref{obe})}}}
 In this section we will prove the one block estimate.
With respect to the usual proofs of similar results, we have here the
difficulty of the boundary conditions, so we need a further averaging
on small macroscopic blocks of length $\varepsilon N$.

We define the space--time average of the distribution
\begin{equation}\label{hat2}
\bar f_{t}^{N,\ell,k} = \frac 1t\int_0^t\frac1{[N(1-2\ell)]}
\sum_{i=N\ell}^{N(1-\ell)} f_{s,i}^{N,k}\left(r_{- k},p_{- k},\dots ,r_{ k},p_{ k}\right)ds 
 \end{equation}
where we defined the projections
\begin{multline}\label{15}
f_{s,i}^{N,k}(\tilde r_{- k},p_{-k}\dots,\tilde r_{ k},\tilde p_{k})\\
=\!\!\int \!\!f_s^N(r_{1},p_1\!,\!\dots\!,\!r_{\!i\!-\! k\!-\!1},p_{i\!-\! k\!-\!1},\tilde r_{\!-
  \!k},\tilde p_{-\! k},\!\dots\!,\!\tilde r_{ k},\tilde p_{ k},r_{\!i+k+1},p_{i+
  k+1}\!,\!\dots\!,\!r_{N},p_{N})\! \!\!\prod_{|i-l|>k}\!\!\!\! d r_l d p_l.
\end{multline}
and we denote
 \begin{equation}\label{hat1}
 d\bar\mu_{t}^{N,\ell,k}:= \bar f_{t}^{N,\ell,k} 
\prod_{|l|\le k} dr_{l}\,dp_{l}
\end{equation}

\subsubsection{{Tightness}\label{Tightness}}

We have the following
\medskip

{\lemma[Tightness]\label{tight} For each $k$ fixed , the sequence  
$(\bar\mu_{t}^{N,\ell,k})_{N\geq1}$ of probability measures is tight.}

\begin{proof}

From the definition of $\bar\mu_{t}^{N,\ell,k}$ we have
\begin{eqnarray*}
&&\int \left(\frac1{2k+1} \sum_{|l|\leq k} e_l \right)\; \bar
f_{t}^{N,\ell,k}\prod_{|l|\leq k} dr_l\;dp_l\\ 
&=&\frac1t \int_0^t \frac1{[N(1-2\ell)]}
\sum_{i=N\ell}^{N(1-\ell)} \left(\int \left(\frac1{2k+1}\sum_{|l|\leq
    k}e_l\right) f_{s,i}^{N,k} \;dr_{-k}\;dp_{-k}\dots
dr_{k}dp_{k}\right)ds\\
 &\le& \frac1t \int_0^t \int
   \left(\frac1{N(1-2\ell)}\sum_{l=1}^N e_l\right) 
   f_{s}^{N} \; \prod_{l\in\mathbb Z} dr_{l}dp_{l} \; ds \le  C
\end{eqnarray*}
by lemma \ref{energybound}, and this implies the tightness.
\end{proof}

Lemma \ref{tight} asserts that for each fixed $k$ there exists a limit
point $\mu_{t}^{\ell,k}$  of the sequence
$(\bar\mu_{t}^{N,\ell,k})_{N\geq1}$.  On the other hand, since
the sequence $(\mu_{t}^{\ell,k})_{k\geq1}$ forms a consistent
family of measures, by Kolmogorov's Theorem, for $k\rightarrow\infty$, 
there exists a unique probability
measure $\mu$ on the configuration space 
$\{(r_i,p_i)_{i\in\mathbb Z}\in (\mathbb R^2)^{\infty}\}$, such that
the restriction of $\mu$ 
on $\{(r_j,p_j)_{j\in\{- k,\dots, + k\}}\in(\mathbb R^2)^{2k+1}\}$ is
$\mu_{t}^{\ell,k}$. 

\subsubsection{\label{Pobe} Proof of the one-block-estimate}

Let us define the formal generator $\mathcal G$ of the infinite dynamics by \index{{\large{CHAPTER 2:}}!$\mathcal G$}
\begin{equation}\label{genZ}
\mathcal G:=\mathcal L+\gamma\mathcal S,
\end{equation}
with the antisymmetric part \index{{\large{CHAPTER 2:}}!$\mathcal L$}
\begin{equation}\label{liouvilleZ}
 \mathcal L:=\sum_{j\in\mathbb Z}\left\{p_j\left(\frac{\partial}{\partial r_{j}}-\frac{\partial}{\partial r_{j+1}}\right)+\left(V^{\prime}(r_{j+1})-V^{\prime}(r_{j})\right)\frac{\partial}{\partial p_j}\right\}
\end{equation}
and the symmetric part \index{{\large{CHAPTER 2:}}!$\mathcal S$}
\begin{equation}\label{symZ}
\mathcal S:= \sum_{i\in\mathbb Z}\left(f\left(\mathbf r,\mathbf p^{j,j+1}\right)-f(\mathbf r,\mathbf p)\right)\\
\end{equation}

In section \ref{proofprop}  we will prove the following Proposition:
{\proposition\label{prop} Any limit point $\mu$ of $\bar
  \mu_{t}^{N,\ell ,k}$, for $N\to\infty$ and then
  $k\to\infty$,  satisfies the following properties:
\begin{enumerate}
\item it has finite entropy density: there exists a constant $C>0$
  such that for all subsets $\Lambda\subset \mathbb Z$  
$$
H\left(\mu|_{\Lambda}\Big|\nu^{|\Lambda|}_{(\tau\beta,0,\beta)}\right)\leq C|\Lambda|,
$$
\item  it is translation invariant: For any local function $F$ and any $j\in\mathbb Z$,
$$
\int F\;d\mu=\int(\tau_j F)\;d\mu
$$ 
where $\tau_j$ denotes the spatial shift by $j$ on the configurations.
\item it is stationary with respect to the operator $\mathcal G$:  For any smooth bounded local function $F$
$$
\int(\mathcal G F)d\mu=0.
$$
\end{enumerate}
}
With this Proposition, we can apply the ergodic theorem from \cite{ffl}:

{\theorem [Ergodicity]\label{char}
Any limit point $\mu$ of 
$\bar \mu_{t}^{N,\ell ,k}(d\mathbf r,d\mathbf p)$  is a convex
combination of Gibbs measures i.e   
$$
\mu(d\mathbf r,d\mathbf p)=\prod_{i\in\mathbb Z}g_{\boldsymbol\lambda}(r_i,p_i)dr_idp_i.
$$
}
The proof of Theorem \ref{char} is contained in \cite{ffl}, see also Chapter 2 of \cite{stefano} for more details. The idea of the proof is the following:
With Proposition \ref{prop} one can prove that $\mu$ is separately
stationary for $\mathcal L$ and $\mathcal S$. This implies that the
distribution of momenta conditioned on position $\mu(d\mathbf
p|\mathbf r)$ is exchangeable. It is here where we use the noise in
the dynamics. For details of this result see \cite{ffl}, or  Chapter
2.3 of  \cite{stefano}. In Chapter 2.2 of the same notes
\cite{stefano} there can be found a 
detailed proof of how this implies \ref{char}. 

\bigskip


\medskip

{\textit {Proof of Theorem \ref{obe}}}:

\medskip
Reacall \eqref{obe0}:
\begin{multline}
\int_0^t \int 
\frac 1N\sum_{i=[N\ell]}^{N-[N\ell]}\left|\frac 1{
    2k+1}\sum_{|j-i|\leq k}\mathbf J^b_{l-1,l}-  
\tilde{\mathbf J}^b\left({\boldsymbol\zeta}_{j}^k\right) 
\right| f_{s}^{N}\, d\mathbf p\,  d\mathbf r\, ds. 
\end{multline}

 By Lemma \ref{tight} and $(ii)$, $(iii)$ of Proposition \ref{prop}, it then is enough to show that for each $b$ and $\ell$ 
$$
\limsup_{k\rightarrow\infty} \sup_{\mu\in\mathcal G}\int\left|\frac1{2k+1}
\sum_{l=-k+1}^k \mathbf J^b_{l-1,l} - \tilde{\mathbf J}^b
\left(\frac1{2k+1} \sum_{l=-k+1}^k \boldsymbol\zeta_{l}\right)\right| d\mu=0
$$
where $\mathcal G$ is the set of Gibbs measures.
But this is just the  law of large numbers and holds in the limit as
$k\rightarrow\infty$.
\begin{flushright}
$\Box$
\end{flushright}

\subsubsection{Proof of Proposition \ref{prop}\label{proofprop}}


{\lemma\label{transinv} Any limit probability $\mu$ of 
$\bar \mu_{t}^{N,\ell,k}$, for $N\to\infty$ and then
  $k\to\infty$,  is  translation invariant.

\begin{proof}

Let $F$ be a bounded, local function depending on configurations
only through $-m,\dots,m$  for some $m\geq 0$. 
Then there exists for each $z\in \mathbb Z$ an integer $k$ such that
$|m+z|\leq k$. 
Since $(\bar f_{t}^{N,\ell,k})_N$ is tight, it suffices to prove
that for each $z$   
\begin{eqnarray}\label{16}
\lim_{k\rightarrow\infty}\lim_{N\rightarrow\infty}\int(F-\tau_z F)
\bar f_{t}^{N,k,\ell} \prod_{|l|\leq k} dr_l d p_l=0 
\end{eqnarray}

The integral  is equal to:
\begin{eqnarray*}
&&\frac 1t\int_0^t\frac1{[N(1-2\ell)]}\sum_{j=N\ell}^{N(1-\ell)}
\int (\tau_j F -\tau_{z+j} F) f_{s,j}^{N,k} \prod_{|l|\leq k} dr_l d p_l\;ds .
\end{eqnarray*}
Because $k\geq|m+z|$, and $F_j$ is bounded,
\begin{eqnarray*}
&&\sum_{j=N\ell}^{N(1-\ell)}
\int(\tau_j F -\tau_{z+j} F) f_{s,j}^{N,k}\prod_{|l|\leq k} dr_l d p_l\\ 
&=&\int \sum_{j=N\ell}^{N(1-\ell)} (\tau_j F -\tau_{z+j} F)
f_s^{N} \;\prod_{r\in\mathbb Z} dr_r dp_r  
=\mathcal O(z),
\end{eqnarray*}
and \eqref{16} follows.

\end{proof}

{\lemma\label{inv} Any limit measure $\mu$ is  stationary in time with
  respect to the generator $\mathcal G=\mathcal L+\gamma\mathcal S$,
  that means for any bounded smooth local function $F(\mathbf
  r,\mathbf p)$ 
\begin{equation}\label{49}
\int \mathcal G F\; d\mu=0.
\end{equation}
}

\begin{proof}
We have to show that for some $k\geq m$

\begin{equation*}
\lim_{k\rightarrow\infty}\lim_{N\rightarrow\infty}\int\mathcal G F_0
\; \bar f_{t}^{N,\ell,k} \prod_{|l|\leq k} dr_l d p_l = 0.
\end{equation*}
With  \eqref{hat1}, the integral is equal to
\begin{eqnarray}\label{50}
 &&\frac 1t\int_0^t\frac1{ [N(1-2\ell)]}\sum_{j= N\ell}^{N(1-\ell)} 
\int\mathcal G F f_{s,j}^{N,k}\;\prod_{|l|\leq k}dr_l d p_l\;ds
\end{eqnarray}
Define the space average 
$$
\bar F:= \frac 1{[N(1-2\ell)]}\sum_{j= N\ell}^{N(1-\ell)}
\tau_j F
$$
Observe that $\mathcal G_N^{\tau(t)}\bar F = \mathcal G\bar F$, 
then we can rewrite  \eqref{50} as
\begin{eqnarray*}
\frac 1{Nt}\int_0^t\int(N\mathcal G_N^{\tau(s)} \bar F)\;
f_s^N\;d\mathbf r\;d\mathbf p\;ds 
&=&\frac 1{tN}\int_0^t \int  \bar F \partial_s f_s^N\; d\mathbf
r\,d\mathbf p ds \\
&=&\frac 1{tN}\left\{\int\bar F f_t^N\; d\mathbf r\, d\mathbf p
-\int \bar F f_0^N\; d\mathbf r\,d\mathbf p\right\}.
\end{eqnarray*}
We conclude the proof by observing that this expression converges to
$0$ if $N\rightarrow\infty$, since $\bar F$ 
is a bounded function. 
\end{proof}
\rm

\subsubsection{Entropy density\label{Cobe}}

For some integer $n\geq1$, define by $\Lambda^n$ a box of length
$2n+1$ and by $\Lambda^n_i$ a box of length $2n+1$ and centered at
$i$. 
Furthermore, let
$$
\nu^{\infty}_{(\tau\beta,0,\beta)}(d\mathbf r,d\mathbf
p):=\prod_{i\in\mathbb Z}\nu_{(\tau\beta,0,\beta)}(dr_i,dp_i)
$$ 
and
$$
H_{\Lambda^{k}}(\mu|\nu^{\infty}_{(\tau\beta,0,\beta)}):=H(\mu|_{\Lambda^k}|\nu^{k}_{(\tau\beta,0,\beta)}).
$$
We obtain the following Lemma:

{\lemma\label{entrdensity} The limit point $\mu$  has finite  entropy
  density, that means there exists a constant $C>0$ such that for all
  subsets $\Lambda^{k}$ 
$$
H_{\Lambda^k}(\mu|\nu^{\infty}_{(\tau\beta,0,\beta)})\leq C|\Lambda^{k}|.
$$
 }
 
 \begin{proof}
By convexity of the relative entropy, we have
\begin{eqnarray}\label{20}
\nonumber&& H\left(\bar \mu_{t}^{N,\ell,k}\; \Big|
  \;\nu^k_{(\tau\beta,0,\beta)}\right)\nonumber
\\
&\leq&\frac1{[N(1-2\ell)]}\sum_{j= N\ell}^{N(1-\ell)} 
H\left(\frac 1t \int_0^t f_{s,j}^{N,k} dr_{-k}dp_{-k}\dots dr_k
  dp_k\;ds\;\Big| \;\nu^k_{(\tau\beta,0,\beta)}\right)\nonumber\\
&=&\frac1{[N(1-2\ell)]}\sum_{j= N\ell}^{N(1-\ell)} 
H_{\Lambda_{j}^{k}}\left(\bar \mu_t^N \;\Big|\;\nu^N_{(\tau\beta,0,\beta)}\right)
\end{eqnarray}
where $\bar\mu_t^N:=\bar f_t^N d\mathbf r\;d\mathbf p$ with
\begin{equation}\label{20bar}
\bar f_t^N:=\frac 1t \int_0^t f_s^N ds.
 \end{equation}
 Relative entropy is superadditive in the following sense (see for
 example \cite{stefano}): let $(\Lambda_i)_{i\in I\subset \mathbb N}$
 be a family of disjoint subsets of $\mathbb Z$. Then 
$$
H_{\bigcup_{i\in I}\Lambda_i}\left(\bar\mu_t^N\;\Big|\;\nu^N_{(\tau\beta,0,\beta)}\right)\geq\sum_{i\in I}H_{\Lambda_i}\left(\bar \mu_t^N\;\Big|\;\nu^N_{(\tau\beta,0,\beta)}\right).
$$
The sum in \eqref{20} can be rearranged in $2k+1$ sums of sums over
disjoint blocks, then applying the superadditivity \eqref{20} is
bounded by 
$$
\frac{(2k+1)}{[N(1-2\ell)]} H\left(\bar \mu_t^N 
  \;\Big|\;\nu^N_{(\tau\beta,0,\beta)}\right).
$$
We will prove in Lemma \ref{entropybound}, that there exists a finite constant $C$ independent of $N$, such that
\begin{equation}\label{ebd}
H\left(\bar \mu_t^N  \;\Big|\;\nu^N_{(\tau\beta,0,\beta)}\right)\leq CN.
\end{equation}
 By Lemma \ref{tight} the sequence
 $(\bar\mu_{t}^{N,\ell,k})_N$ is tight. Since by Lemma
 \ref{transinv} each limit point $\mu$ of
 $(\bar\mu_{t}^{N,\ell,k})_N$  is translation invariant and
 stationary, we can conclude the proof with  the lower semicontinuity
 of the relative entropy.  
\end{proof}

{\rm To complete the proof of Lemma \ref {entrdensity} it remains to show
\eqref{ebd}.  }

{\lemma\label{entropybound} 
  If 
  $$
  H\left(f_0^Nd\mathbf rd\mathbf p\,|\,\nu^N_{(\tau\beta,0,\beta)}\right)\leq C_1N
  $$
 for some uniform constant  $C_1>0$, then for any $N\in\mathbb N$
 there exists a constant $C_2>0$ such that
 $$
H\left(\bar \mu_t^N \,|\,\nu^N_{(\tau\beta,0,\beta)}\right)\leq C_2N,
$$ 
where $\bar \mu_t^N$ is defined by \eqref{20bar}.
}

\begin{proof}
Recall from Lemma \ref{dineq}, that the relative entropy with respect
to Lebesgue measure is nonincreasing in time since the Lebesgue
measure is stationary with respect to the generator $\mathcal
G_N^{\tau(t)}$ (for any $t$). 
Therefore, as in the  proof of Lemma \ref{dineq} we can write  
\begin{eqnarray*}
&&H\left(f_{t}^Nd\mathbf r\,d\mathbf
  p\,\big|\,g_{(\tau\beta,0,\beta)}^N\,d\mathbf r\,d\mathbf
  p\right)-H\left(f_{0}^N d\mathbf r\,d\mathbf p\,\big|\,g_{(\tau\beta,0,\beta)}^Nd\mathbf r\,d\mathbf p\right)\\
&\leq &-\int \log g_{(\tau\beta,0,\beta)}^N f_{t}^N 
d\mathbf r\,d\mathbf p 
+\int \log g_{(\tau\beta,0,\beta)}^N f_{0}^N d\mathbf r\,d\mathbf p.
\end{eqnarray*}
The  last line is  then equal to
\begin{eqnarray*}
&=&-\int_0^t\int f_s^N NG_N^{\tau(s)}\log g_{(\tau\beta,0,\beta)}^N  d\mathbf r\,d\mathbf p ds\\
&=& - \beta N \int_0^t \tau(s) \int f_s^N  p_N d\mathbf r\,d\mathbf p\; ds.
\end{eqnarray*}
Since the last line is equal to the expectation of $\beta \sum_{j=1}^N
(e_j(t) - e_j(0))$, by lemma \ref{energybound} it is bounded by $CN$
for some constant $C$.

Hence, by convexity of $H(\cdot|\cdot)$,
$$
H\left(\bar \mu_t^N \,|\,\nu^N_{(\tau\beta,0,\beta)}\right)\leq (C_1+C_3)N.
$$ 
\end{proof}

{\lemma\label{energybound} 
  If the initial configuration satisfy
  $$
  \sum_{j=1}^N e_j(0) \le CN 
  $$
then there exists a constant $\tilde C(t)$ independent of N such that

\begin{equation}
 \sum_{j=1}^N e_j(t) \le \tilde C(t) N\label{eq:enbound}
\end{equation}

}

\begin{proof}
  Define
  \begin{equation*}
    F_N(t) = \sum_{j=1}^N e_j(t) - \tau(t) q_N(t) = \sum_{j=1}^N 
    \left(e_j(t) - \tau(t) r_j(t)\right) 
  \end{equation*}
Computing the time evolution of this function we have
\begin{equation}\label{Fevol}
  F_N(t) = F_N(0) - \int_0^t \tau'(s) q_N(s) \; ds
\end{equation}
Consequently 
\begin{equation}
  \label{eq:2}
   \sum_{j=1}^N e_j(t) = \tau(t) q_N(t) - \tau(0) q_N(0) +
   \sum_{j=1}^N e_j(0) - \int_0^t \tau'(s) q_N(s) \; ds
\end{equation}
By condition \eqref{V1}, we have that 
$$
|q_N|\le \sum_j |r_j| \le \sqrt N \left(\sum_j |r_j|^2\right)^{1/2} \le
  C_{-}^{-1/2}\sqrt N  \left(\sum_j V(r_j) \right)^{1/2} \le
  C_{-}^{-1/2}\sqrt N  \left(\sum_j e_j \right)^{1/2} 
$$
Then we can estimate
\begin{equation*}
  \begin{split}
    \left| \int_0^t \tau'(s) q_N(s) \; ds\right| \le
    \|\tau'\|_\infty \int_0^t \left| q_N(s)\right| ds 
    \le  \frac{\|\tau'\|_\infty}{C_-^{1/2}} \sqrt N \int_0^t \left(\sum_{j=1}^N
      e_j(s)\right)^{1/2} ds
  \end{split}
\end{equation*}
Defining $\bar e_N(t) = \frac 1N \sum_{j=1}^N e_j(t)$ we have then
\begin{equation*}
  \bar e_N(t) \le \|\tau\|_\infty \left(\sqrt{\bar e_N(t)} +
    \sqrt{\bar e_N(0)}\right) + \bar e_N(0) +
  \frac{\|\tau'\|_\infty}{C_-^{1/2}} \int_0^t
  \sqrt{\bar e_N(s)} ds
\end{equation*}
that implies \eqref{eq:enbound}.
\end{proof}


\bibliographystyle{amsalpha}

{Nadine Braxmeier-Even \\
Institut f{\"u}r Mathematik\\
Universit{\"a}t W{\"u}rzburg\\
Campus Hubland Nord\\
Emil-Fischer-Stra{\ss}e 30\\
97074 W{\"u}rzburg, Germany
\\
\texttt{{nadine.braxmeier-even@mathematik.uni-wuerzburg.de}}
}
\\
\\
{Stefano Olla \\
CEREMADE, UMR CNRS 7534\\
Universit\'e Paris-Dauphine\\
75775 Paris-Cedex 16, France, \\
\texttt{{ olla@ceremade.dauphine.fr}}
}

\end{document}